\documentclass{elsart}

\usepackage{amssymb}

\def\be{\begin{equation}}
\def\ee{\end{equation}}
\def\bea{\begin{eqnarray}}
\def\eea{\end{eqnarray}}

\newcommand{\nn}{\nonumber}

\newcommand\R{\mathbb{R}}
\newcommand\Z{\mathbb{Z}}

\newcommand\C{\mathbb{C}}
\newcommand\T{\mathbb{T}}
\newcommand\diff{\mathrm{d}}
\newcommand{\de}{\partial}

\begin{document}

\begin{frontmatter}

{\small\hfill CERN-PH-TH/2005-081}\\
{\small\hfill HUTP-05/A0025}\\



\title{Toric Sasaki--Einstein metrics on $S^2\times S^3$}
\author{Dario Martelli}
\address{Department of Physics, CERN Theory Division\\
1211 Geneva 23, Switzerland}
\ead{dario.martelli@cern.ch}

\author{James Sparks}
\address{Department of Mathematics, Harvard University \\
One Oxford Street, Cambridge, MA 02318, U.S.A.}
\address{Jefferson Physical Laboratory, Harvard University \\
Cambridge, MA 02138, U.S.A.}
\ead{sparks@math.harvard.edu}

\begin{abstract}
We show that by taking a certain scaling limit of a Euclideanised 
form of the Plebanski--Demianski metrics one obtains a family of local 
toric K\"ahler--Einstein metrics. These can be used to construct
local Sasaki--Einstein metrics in five dimensions which are generalisations
of the $Y^{p,q}$ manifolds. In fact, we find that these  
metrics are diffeomorphic to those recently found by Cvetic, Lu, Page and
Pope. We argue that the corresponding family of  
smooth Sasaki--Einstein manifolds all have topology 
$S^2 \times S^3$. We conclude by setting up the equations describing the warped
version of the Calabi--Yau cones, supporting $(2,1)$ three--form flux.
\end{abstract}

\end{frontmatter}

Recently Sasaki--Einstein geometry has been the focus of much attention. 
The interest in this subject has arisen due to the discovery
in \cite{paper1,paper2} of an infinite family $Y^{p,q}$ of 
explicit Sasaki--Einstein metrics on $S^2\times S^3$, and the subsequent 
identification of the corresponding family of AdS/CFT dual quiver 
gauge theories in \cite{toric,quivers}. 

The construction of \cite{paper2} was immediately generalised to higher 
dimension in reference \cite{paper3} and a further generalisation 
subsequently appeared in \cite{strasbourg,morese}. However, dimension 
five is the most interesting dimension physically and the purpose 
of this work was to investigate if there exist other local 
K\"ahler--Einstein metrics in dimension four from which one can 
construct complete Sasaki--Einstein manifolds in one dimension higher. 

As we show, one can obtain a family of local toric K\"ahler--Einstein 
metrics by taking a certain scaling limit of a Euclideanised 
form of the Plebanski--Demianski metrics \cite{PD}. Here toric 
refers to the fact that the metric has two commuting holomorphic 
Killing vector fields. In fact the resulting metrics were found independently by 
Apostolov and collaborators in \cite{oops} using rather different methods. 
In the latter reference 
it is shown that this family of metrics constitute the most general local 
K\"ahler--Einstein metric which 
is \emph{orthotoric}, a term that we define later. 
We also show that these K\"ahler--Einstein metrics are precisely 
those used in the recent construction of Sasaki--Einstein manifolds generalising 
$Y^{p,q}$ \cite{cvetic}. Higher--dimensional 
orthotoric K\"ahler--Einstein 
metrics are given in explicit form in reference \cite{apostolov}.

Our starting point will be the following family of local Einstein metrics 
in dimension four
%
\be 
\diff s^2_4 = (p^2 - q^2) \left[ \frac{\diff p^2}{P} + \frac{\diff 
q^2}{Q} \right]+ \frac{1}{p^2-q^2}
\left[P(\diff \tau + q^2 \diff \sigma )^2 +
Q(\diff \tau + p^2 \diff \sigma )^2\right]
\label{PDmet}
\ee
where $P$ and $Q$ are the fourth order polynomials
\begin{eqnarray} 
P (p)&=& - \kappa (p-r_1)(p-r_2)(p-r_3)(p-r_4) \,\nn\\ 
Q (q)&=&  \kappa (q-r_1)(q-r_2)(q-r_3)(q-r_4) +cq\,,\\ 
0 & = & r_1 + r_2 +r_3 +r_4~.\label{cons}
\end{eqnarray}
As shown in \cite{quater}, these metrics arise by taking a scaling limit 
of the well--known Plebanski--Demianski metrics \cite{PD}. 
The Weyl tensor is anti--self--dual if and only if $c=0$. 

The natural almost K\"ahler two--form associated to the metric (\ref{PDmet}) is
\be
J = \diff p \wedge (\diff\tau +q^2 \diff\sigma) + 
\diff q \wedge (\diff \tau +p^2 \diff\sigma) ~.
\label{tform}
\ee
Our strategy will be to obtain a scaling limit for 
which (\ref{tform}) becomes closed. Thus, consider the following 
change of coordinates 
\bea
p & =  1- \epsilon \xi \qquad  & q = 1- \epsilon \eta\nn\\
\Phi & =  \epsilon (\tau + \sigma )\qquad & \Psi  =  -2\epsilon^2 \sigma
\eea
and redefinition of the metric constants
\bea
 r_i & = & 1- \epsilon \alpha_i\qquad i=1,2,3\nn\\
 r_4 & = & -3 +(\alpha_1+\alpha_2+\alpha_3) \epsilon\nn\\
c & = & 4\epsilon^3\gamma~. 
\eea
The latter ensures that the constraint (\ref{cons}) is satisfied. Defining
\bea
F(\xi) & = &-\kappa(\alpha_1-\xi)(\alpha_2-\xi)(\alpha_3-\xi)\nn\\
G(\eta) & = & \kappa(\alpha_1-\eta)(\alpha_2-\eta)(\alpha_3-\eta)+\gamma
\eea
it is straightforward to see that, upon sending $\epsilon \to 0 $, the metric
(\ref{PDmet}) becomes
\bea
\diff s_4^2 =  \frac{(\eta-\xi)}{2F(\xi)}\diff\xi^2 + \frac{2F(\xi)}{(\eta-\xi)} (\diff\Phi
+\eta\diff\Psi)^2
+ \frac{(\eta-\xi)}{2G(\eta)}\diff \eta^2 + 
\frac{2G(\eta)}{(\eta-\xi)} (\diff\Phi +\xi\diff \Psi)^2~. \nn\\
\label{newkahler}
\eea
In fact it is also immediate to see that $J=2\diff A$ where
\be
-A = \frac{1}{2} (\xi+\eta) \diff\Phi +\frac{1}{2} \xi\eta \diff\Psi~.
\ee
One can verify that this metric is K\"ahler--Einstein with 
curvature 
\bea
\mathrm{Ric} =  3\kappa g~.
\eea
In particular, setting $\kappa=2$ the metric 
\be\label{semetric}
\diff s^2_5 = \diff s^2_4 + (\diff\psi' + A )^2
\label{loca}
\ee
is then locally Sasaki--Einstein with curvature $4$. (See \emph{e.g.} \cite{paper3}
for curvature conventions).

Having found these metrics we subsequently discovered 
that the same solutions had been 
obtained independently, and in a completely different manner, 
in reference \cite{oops}. In fact the metric, 
as presented, is essentially already in the form given in \cite{oops} 
and moreover is the most general orthotoric 
K\"ahler--Einstein metric. Here orthotoric means that 
the Hamiltonian functions $\xi+\eta$, $\xi\eta$ for the Killing vector 
fields $\de/\de\Phi$, $\de/\de\Psi$, respectively, have the property that 
the one--forms $\diff\xi$, $\diff\eta$ are orthogonal. 
The metric is self--dual if and only if 
$\gamma=0$. Moreover these metrics have been generalised to 
arbitrary dimension in \cite{apostolov} which thus gives a more
general construction of local Sasaki--Einstein metrics.

One should now proceed to analyse when the local Sasaki--Einstein metrics 
extend to complete metrics on a smooth manifold. 
As the metrics generically possess three commuting Killing
vectors $\de/\de\Phi, \de/\de \Psi, \de /\de \psi'$, the resulting 
five dimensional 
manifolds should be toric, as was the case in 
\cite{paper2}. In particular, real 
codimension two fixed point sets correspond to toric divisors \cite{toric} 
in the Calabi--Yau cone, and 
it is a very simple matter to find such vector fields 
for the metric (\ref{semetric}).

A generic Killing vector can be written as
\bea
V & = & S \frac{\de}{\de \Phi} + T \frac{\de}{\de \Psi}+U \frac{\de}{\de \psi'}
\eea
where $S,T$ and $U$ are constants. A short calculation then shows that its norm is given by
\bea
||V||^2  =  \frac{2}{(\eta-\xi)}\left[ F(\xi) (S+T\eta )^2 + G(\eta) 
(S+T\xi)^2\right]\nn\\ 
+\frac{1}{4} \left[S(\xi+\eta)+T\xi\eta-2U\right]^2~.
\eea
Now, crucially, since $F(\xi)/(\eta-\xi)>0, G(\eta)/( \eta-\xi)>0$ for 
a positive definite metric, this is a sum of positive 
functions. Therefore it can vanish in codimension two 
if and only if $\xi=\xi_i$ (or $\eta=\eta_i$) are at the roots
of $F$ (or $G$) and at the same time the remaining terms manage to vanish for generic values
of  $\eta$ (or $\xi$). In fact, it is easy to see that this is true.
We therefore see that there are \emph{four} codimension two fixed point sets, so that if the
metrics extend onto complete smooth toric manifolds, 
the Calabi--Yau cones must be a $\mathbb{T}^3$
fibration over a four faceted polyhedral cone in $\mathbb{R}^3$.

However, we will not complete the details 
of this argument because it turns out that these  
metrics are diffeomorphic to those found by Cvetic, Lu, Page, and Pope in \cite{cvetic}.
These authors have performed the global analysis in detail. Therefore, 
we instead exhibit an explicit change of coordinates, 
demonstrating the equivalence of the two metrics.

The K\"ahler--Einstein metrics in reference \cite{cvetic} were given 
in the form
\bea\label{popes}
\diff s^2_4 & = & \frac{\rho^2\diff x^2}{4\Delta_x}+\frac{\rho^2\diff\theta^2}
{\Delta_{\theta}} + \frac{\Delta_x}{\rho^2}\left(\frac{\sin^2\theta}{\alpha}
\diff\phi+\frac{\cos^2\theta}{\beta}\diff\psi\right)^2\nn\\
&& + \frac{\Delta_{\theta}\sin^2\theta\cos^2\theta}{\rho^2}\left(
\frac{\alpha-x}{\alpha}\diff \phi-\frac{\beta-x}{\beta}\diff\psi\right)^2\eea
where
\bea
\Delta_x & = & x(\alpha-x)(\beta-x)-\mu,\qquad \rho^2=\Delta_{\theta}-x\nn\\
\Delta_{\theta} & = & \alpha\cos^2\theta +\beta\sin^2\theta~.\eea
Consider the coordinate transformation\footnote{Note that this is degenerate
when $\alpha=\beta$, which corresponds to the $Y^{p,q}$ limit \cite{cvetic}.}
\bea
\eta & = & \alpha - x, \qquad \xi = (\alpha-\beta)\sin^2\theta\nn\\
\Phi  & = & \frac{1}{2\beta}\psi, \qquad \Psi = \frac{\phi}{2(\alpha-\beta)\alpha} - 
\frac{\psi}{2(\alpha-\beta)\beta}~.
\eea
It is a simple exercise to show that, in these coordinates, 
the metric (\ref{popes}) 
takes the form (\ref{newkahler}) where $\alpha,\beta$ and $\mu$ 
parametrise the cubic function. Explicitly, we have
\bea
F(\xi) & = & 2 \xi (\alpha-\xi) (\alpha-\beta -\xi )\nn\\
G(\eta) & = & -2 \eta (\alpha-\eta) (\alpha-\beta -\eta )-2\mu~.
\eea

As shown in \cite{cvetic}, the complete metrics $L^{a,b,c}$ with local form
(\ref{semetric}) are specified by three integers $a,b,c$. One recovers 
the $Y^{p,q}$ metrics in the limit $a=p-q,b=p+q,c=p$.
Moreover, as explained, there
are precisely four Killing vector fields $V_i$, $i=1,2,3,4$, that vanish on codimension 2
submanifolds. This means that the image of the Calabi--Yau cone under the 
moment map for the $\T^3$ action is a four faceted polyhedral cone in $\R^3$ 
-- see \cite{toric} for a review. Indeed, using the linear relation among the 
vectors in 
\cite{cvetic} one can show that the normal vectors to this polyhedral cone 
satisfy the relation
\be
av_1+bv_2-cv_3-(a+b-c)v_4=0\ee
where $v_i$, $i=1,2,3,4$ are the primitive vectors in $\R^3$ that define 
the cone. From the Delzant theorem 
in \cite{L} it follows that, for $a,b,c$ relatively prime, the Sasaki--Einstein metrics $L^{a,b,c}$ are equivariantly 
contactomorphic to the link of the symplectic quotient
\be\label{symplectic}
\C^4//(a,b,-c,-a-b+c)~.\ee
Note that indeed in the $Y^{p,q}$ limit we obtain charges $(p-q,p+q,-p,-p)$ 
which is the result of \cite{toric}. 
The base $Y$ of the cone is non--singular if $a$ and $b$ are pairwise prime to each of $c$ 
and $a+b-c$, and these integers are strictly positive. 
By the results of \cite{Ltop} we then have $\pi_2(Y)\cong\Z$. Since 
$Y$ is simply--connected, spin and has no torsion in $H_2(Y)$ it follows from 
Smale's Theorem that $Y$ is diffeomorphic to 
$S^2\times S^3$.

We conclude with some technical computations that may be of use in future
developments. In particular,  following \cite{toric}, we first introduce 
complex coordinates on the Calabi--Yau cone. We 
then write down the equations for a warped
version of the Calabi--Yau cone, thus generalising the solution of
\cite{klebanov}. Note that the form of the metric that we have 
presented here is more
symmetric than in the coordinate system of \cite{cvetic}. 

The holomorphic $(3,0)$ form on the cone can be written in the 
standard fashion
\bea
\Omega & = & e^{i\psi'} r^2 \Omega_4 
\wedge \left[ \diff r + i r (\diff \psi' +A)\right]
\eea
with appropriate $\Omega_4$ \cite{toric}. Introducing the one--forms
\bea
\hat\eta_1 & = & \frac{1}{2F}\diff \xi + 
\frac{i}{\eta-\xi} (\diff \Phi + \eta \diff \Psi)\nn\\
\hat\eta_2 & = & \frac{1}{2G}\diff \eta + 
\frac{i}{\eta-\xi} (\diff \Phi + \xi \diff \Psi)\nn\\
\hat\eta_3 & = & \frac{\diff r}{r}+ i  (\diff \psi' +A)
\eea
this may be written as
\bea
\Omega & = & 2 (\eta -\xi )\sqrt{FG} r^3 e^{i\psi'}\, \hat \eta_1 \wedge  
\hat \eta_2 \wedge  \hat \eta_3 ~.
\label{cunta}
\eea
By construction the $\hat \eta_i$ are such that $\hat \eta_i \wedge \Omega=0$, 
\emph{i.e.} they  are $(1,0)$ forms. However, one must take combinations of these to
obtain integrable (closed) forms. These are given by 
\bea
\eta_1 & = & \hat \eta_1 - \hat \eta_2 = \frac{1}{2F}\diff \xi - \frac{1}{2G} \diff \eta
+ i \diff \Psi\nn\\
\eta_2 & = & \xi \, \hat \eta_1 - \eta \, \hat \eta_2 =
\frac{\xi}{2F}\diff \xi - \frac{\eta}{2G} \diff \eta - i \diff \Phi\nn\\
\eta_3 & = & \xi^2 \, \hat \eta_1 - \eta^2 \, \hat \eta_2 - 2 \hat\eta_3 =
\frac{\xi^2}{2F}\diff \xi - \frac{\eta^2}{2G} \diff \eta -2 \frac{\diff r }{r}
- 2 i \diff \psi'~.
\eea
The effect of this change of basis is to simplify (\ref{cunta}) slightly 
\bea
\Omega & = & \sqrt{FG} r^3 e^{i\psi'}\, \eta_1 \wedge  
\eta_2 \wedge  \eta_3 ~.
\eea
Integrating these one--forms, thus introducing $\eta_i = \diff z_i /z_i$, 
we obtain the following set of complex coordinates:
\bea
z_1 & = & \prod_{i=1}^3 (\xi - \xi_i)^{\frac{1}{4\prod_{j\neq i} 
(\xi_i-\xi_j)}}(\eta - \eta_i)^{\frac{1}{4\prod_{j\neq i} 
(\eta_i-\eta_j)}} e^{i\Psi}\nn\\
z_2 & = &  \prod_{i=1}^3 (\xi - \xi_i)^{\frac{\xi_i}{4\prod_{j\neq i} 
(\xi_i-\xi_j)}}(\eta - \eta_i)^{\frac{\eta_i}{4\prod_{j\neq i} 
(\eta_i-\eta_j)}} e^{-i\Phi}\nn\\
z_2 & = &  \prod_{i=1}^3 (\xi - \xi_i)^{\frac{\xi_i^2}{4\prod_{j\neq i} 
(\xi_i-\xi_j)}}(\eta - \eta_i)^{\frac{\eta_i^2}{4\prod_{j\neq i} 
(\eta_i-\eta_j)}} r^{-2}e^{-2i{\psi}^{\prime}}
\eea
For convenience we have written the cubic polynomials as 
\bea
F(\xi)  & = & 2(\xi - \xi_1) (\xi -\xi_2) (\xi -\xi_3)\nn\\
G(\eta) & = & -2(\eta - \eta_1) (\eta -\eta_2) (\eta -\eta_3)~.
\eea
These coordinates generalise those introduced in \cite{toric} for the 
metric cone over the $Y^{p,q}$ manifolds.

Next we turn to  the problem of finding warped solutions which arise
after placing fractional branes at the apex of the cone. These solutions are
expected to be relevant for the study of cascades in the dual gauge theories.
In the following we follow the logic of reference \cite{klebanov}. 
Recall that in Type IIB supergravity 
it is possible to turn on a complex three--form flux preserving supersymmetry,
provided this is of Hodge type $(2,1)$ with respect to the complex structure of
the Calabi--Yau cone. Such a three--form is easily constructed in terms of a
local closed primitive $(1,1)$ form on the K\"ahler--Einstein space. It is
straightforward to see that such a two--form is
\bea
\omega = \frac{1}{(\xi-\eta)^2}\left[ \diff (\xi-\eta)\wedge \diff \Phi + 
(\eta \diff \xi - \xi \diff \eta)\wedge \diff \Psi\right]~.
\eea
Then, by construction 
\bea
\Omega_{2,1} = \left[\frac{\diff r}{r} + i (\diff \psi'+A)\right] \wedge \omega
\eea
is $(2,1)$ and closed. 
Next, let us give the scalar Laplacian operator on the Calabi--Yau
cone, acting on $\mathbb{T}^3$--invariant functions. Again, this is highly 
symmetric in $\xi, \eta$ due to the particularly simple
form of the metric:
\bea
\Delta_{CY} = \frac{1}{r^5(\eta-\xi)}\left[ (\eta-\xi) \de_r (r^5\de_r)
+2r^3\left(\de_{\eta} (G\de_{\eta}) + \de_\xi (F\de_\xi)\right)\right]~.
\eea
One is then interested in finding solutions of the type
\bea
\diff s^2 &=& h^{-1/2} \diff s^2 (\mathbb{R}^4) + 
h^{1/2} (\diff r^2 + r^2 \diff s^2_5)\nn\\
 F_3 + i H_3  & \propto  &\Omega_{2,1} ~,
\eea
where $F_3$ and $H_3$ are the RR and NS three--forms respectively,
\cite{klebanov},
for which the only non--trivial equation reduces to 
\bea
\Delta_{CY} h = -\frac{1}{6} |H_3|^2~.
\eea
After the substitutions \cite{klebanov}
\be
h (r,\xi,\eta)= r^{-4} \left[\frac{A}{2} t + s(\xi,\eta) \right]~,\qquad \quad
t = \log r
\ee
we are left with the following PDE
\bea
\frac{\de}{\de \eta } \Big(G(\eta)\frac{\de}{\de \eta} s(\xi,\eta)\Big)+ 
\frac{\de}{\de \xi } \Big(F(\xi)\frac{\de}{\de \xi} s(\xi,\eta)\Big)=
-\frac{C^2}{(\xi-\eta)^3}+ A (\xi -\eta) 
\eea
where $C$ is a proportionality constant. Of course this is still dependent 
on two variables, 
but it seems that  solutions generalising those of \cite{klebanov}
should exist.

\subsection*{Acknowledgments}
\noindent 
J. F. S. is supported by NSF grants DMS--0244464, DMS--0074329 and DMS--9803347.



\end{document}